\documentstyle[12pt]{article}
\begin{document}
\large
\begin{center}
\bf{An Upper Mass Bound of a Two-Component Protogalaxy}\\
\vspace{15pt}
\normalsize
\rm
    Keiko {\sc Miyahata} and Satoru {\sc Ikeuchi} \\
\vspace{1cm}
\it
 Department of Earth and Space Science, Faculty of Science,
 Osaka University,\\
 Machikaneyama-cho, Toyonaka, Osaka 560
\vspace{12pt}\\
 E-mail miyahata@vega.ess.sci.osaka-u.ac.jp,
ikeuchi@vega.ess.sci.osaka-u.ac.jp
\vspace{0.5cm}\\
\rm
(Received 1995 April 24; accepted 1995 September 18)
\vspace{1.0cm}\\
\end{center}
\sloppy{
\begin{center}
\large
\bf{Abstract} \\
\end{center}
\normalsize

\baselineskip 8mm

{}~~~We investigate the physical properties of a two-component virialized
protogalaxy comprising a hot gas and cold clouds which are in pressure
equilibrium under the assumptions that the protogalaxy is both
spherical and homogeneous.
Two conditions which we adopt are that the protogalaxy is in virial
equilibrium and that the cooling time is less than the dynamical time.
It is found that there exists an upper mass bound for such a two-component
 protogalaxy when the cloud mass is comparable to or greater than that of
hot gas.
We also consider the stability of the cold cloud against
the Jeans instability and Kelvin-Helmholtz instability.\\

\vspace{1.5cm}\\
\large
\bf{1.~Introduction}\\
\rm
\normalsize

\baselineskip 8mm

{}~~~Recently, a remarkable series of observations has revealed many
high-redshift
galaxies, radio sources, and quasars to the redshifts greater than four,
thus suggesting important implications about galaxy formation and
evolution (McCarthy 1993).
Theoretically, from the 1980s many large simulations
concerning galaxy formation in the context of large-scale
structure formation
have been performed (e.g. Efstatiou, Silk 1983; Cen, Ostriker
1992 and references there in).
Since there still remain many uncertain processes in galaxy formation,
it could be valuable to clarify the fundamental nature of galaxy formation
by a simple analytic treatment.

A few decades ago several attempts were proposed (Ostriker 1974, talk at
 7th Texas Conference; Gott,
Thuan 1976; Rees, Ostriker 1977; Silk 1977; Rees 1978). In these studies,
 it was proposed that a protogalaxy consists of one component homogeneous
gas in a virial equilibrium state;
it was assumed that radiative cooling controls
the fate of the protogalaxy, with the ratio of cooling time to the
 collapse time being a controlling parameter. Their conclusions are that
if free-free cooling dominates over bound-free cooling, the upper
bound to the radius of a protogalaxy can be
uniquely expressed by only
 fundamental constants, and that if bound-free cooling predominates, its mass
is uniquely expressed as well.

On the othe hand, another model was proposed in which
the protogalaxy comprises
a hot gas and cold clouds (Fall, Rees 1985) because the hot gas component
is thermally unstable to make cold clouds with mass around globular clusters.
Regarding  such a two-component model of a
protogalaxy, Ikeuchi and Norman
(1991) examined the equilibrium structure, and obtained the expressions
 of its physical quantities by using fundamental constants and several
additional parameters.
Recently, Mathews and Schramn (1993) and Lee et al. (1994) have studied the
star-formation history of a two-component galactic
 halo within the expanding background universe. They emphasized that in order
to explain
the age discrepancy in the galaxy two different types of star-formation
modes are required.
One is such an efficient star-formation as induced by
cloud-cloud collisions
 in the halo; the other is the quiescent star-formation in the disk.

In this paper we consider the physical properties of a two-component
virialized protogalaxy by a simple analysis, and give an upper mass bound
for it when the cloud mass dominates over the hot gas mass.
We also consider the stability of cold clouds against
the Jeans instability and Kelvin-Helmholtz instability.
\vspace{1.0cm}\\

\large
\bf{2. An Upper Mass Bound}
\vspace{0.5cm}\\
\normalsize
\it
{2.1. Basic Equations and Assumptions}
\rm
\baselineskip 8mm

Following the discussion by Ikeuchi and Norman (1991),
we assume for simplicity that this two-component protogalaxy
is spherical and homogeneous.
Suppose that $N_{\rm{c}}$ cold clouds of each radius $R_{\rm{c}}$, mass
$M_{\rm{c}}$, and temperature
$T_{\rm{c}}$ are embedded in a hot gas of radius $R_{\rm{h}}$,
mass $M_{\rm{h}}$, and temperature $T_{\rm{h}}$.
To determine the physical properties of this two-component protogalaxy,
seven equations, or reasonable assumptions
for seven quantities in the above, are needed.
 They are as follows:
\\
(i) Virial equilibrium,
\begin{eqnarray}
 \sigma_{\rm{v}}^{\rm2} &=& 0.6\frac{G(M_{\rm{h}}+N_{\rm{c}}M_{\rm{c})}}
                           {R_{\rm{h}}}
                    = 3\frac{kT_{\rm{h}}}{{\mu}_{\rm{h}}m_{\rm{p}}},
\label{eqn:vir}
\end{eqnarray}
where $\sigma_{\rm{v}}(T_{\rm{h}}),\mu_{\rm{h}}$, and $m_{\rm{p}}$ are
the virial velocity of
protogalaxy, the mean molecular weight of hot ambient gas, and the
proton mass, respectively.
For a primordial wholly ionized gas, $\mu_{\rm{h}}=0.609$.
 We do not explicitly include the
gravity  of dark matter in this paper.\\
(ii) Equality of the free-fall time and cooling time
($\tau_{\rm{ff}}=\tau_{\rm{cool}}$),
\begin{eqnarray}
  \sqrt{\frac{3\pi}{32}}
  \sqrt{\frac{c_{\rm{v}}R_{\rm{h}}^{3}}{G(M_{\rm{h}}+N_{\rm{c}}M_{\rm{c})}}}
       &=& 1.5 kT_{\rm{h}}
      \frac
      {c_{\rm{v}}(R_{\rm{h}}^{3}-N_{\rm{c}}R_{\rm{c}}^{3})
      \mu_{\rm{h}}m_{\rm{p}}}
      {M_{\rm{h}}c_{\rm{\Lambda}}\Lambda(T_{\rm{h}})},
\label{eqn:tim}
\end{eqnarray}
where $c_{\rm{v}}=4{\pi}/3$ is a volume element,
 $\Lambda(T_{\rm{h}})$ is the cooling function of a primordial
 gas in units of $\rm{erg~s^{-1}~cm^{3}}$~(Binney, Tremaine 1987)
and $c_{\rm{\Lambda}}=0.821$ is a
correction factor to the fitting formulae for the cooling function.
Strictly speaking, the protogalaxy will collapse when
the condition $\tau_{\rm{ff}}>\tau_{\rm{cool}}$ is satisfied.
We should thus note that equation
(\ref{eqn:tim}) gives the maximum mass/minimum radius of a collapsing
protogalaxy. If the luminous mass of the galaxy is condensed in the dark halo
 (White, Rees 1978), we must consider a vast amount of
dark matter. A zeroth-order incorporation of this effect can be achieved by
changing the gravitational constant G to G[1+$M_{\rm{d}}/M_{\rm{b}}$]
in equations (\ref{eqn:vir}) and
(\ref{eqn:tim}), where $M_{\rm{d}}$ and $M_{\rm{b}}$ are the masses
of the dark
matter and baryonic matter, respectively.\\
(iii) Pressure equilibrium between clouds and hot ambient gas,
\begin{eqnarray}
 \frac{\rho_{\rm{h}}}{\mu_{\rm{h}}m_{\rm{p}}}T_{\rm{h}} &=&
 \frac{\rho_{\rm{c}}}{\mu_{\rm{c}}m_{\rm{p}}}T_{\rm{c}},
\label{eqn:pre}
\end{eqnarray}
where $\mu_{\rm{c}}$ is the mean molecular weight of the gas in cold
clouds of primordial abundance,
$\mu_{\rm{c}}=$1.23.\\
(iv) Energy balance condition,
\begin{eqnarray}
 n_{\rm{h}}^{2}c_{\rm{\Lambda}}^{2}\Lambda(T_{\rm{h}}) c_{\rm{v}}
 (R_{\rm{h}}^{3}-N_{\rm{c}}R_{\rm{c}}^{3})
&=& \frac{N_{\rm{c}}}{c_{\rm{v}}R_{\rm{h}}^{3}}\pi R_{\rm{c}}^{2}
   \sigma_{\rm{v}}
    \eta(N_{\rm{c}}M_{\rm{c}}c^{2}).
\label{eqn:ene}
\end{eqnarray}
Here, we assume that the
energy loss from the hot region is supplied by
the energy liberation by the cloud-cloud collisions,
and $\eta$ denotes the fraction
of the liberated energy to the cloud rest mass. In the following, we suppose
 that $\eta$ ${\approx}$ $10^{-5}$---$10^{-6}$.
These values are the observed ones in molecular clouds,
considering that the mass fraction of energy
liberation from newly born stars is $\sim 10^{-3}$,
 and the star-formation efficiency from clouds is in the order of
$10^{-2}$---$10^{-3}$.
We suppose that this cloud-cloud collision induces efficient star-formation
during an early stage of protogalactic evolution.
For our model to be consistent, the cold
clouds must be stable (discussed in section 3).

Conditions (i)---(iv) are the physical assumptions for a two-phase
system.
To determine seven quantities, we need to specify additional three conditions,
 though we do not have appropriate ones at present. We regard the three
quantities $T_{\rm{h}},T_{\rm{c}}$ and $F_{\rm{v}}$ as being free parameters,
where $F_{\rm{v}}$ is
the volume-filling factor of clouds,
\begin{eqnarray}
F_{\rm{v}}=\frac{N_{\rm{c}}R_{\rm{c}}^{3}}{R_{\rm{h}}^{3}}.
\end{eqnarray}
Using the characteristics of the cooling function of primordial gas,
we suppose that the hot gas temperature $T_{\rm{h}}$ is $>10^{6.26}$K,
in which
the hot gas is thermally stable,
and that the cold clouds temperature $T_{\rm{c}}$ is $10^{\rm{4}}$K,
in which the primordial cooling function has a sharp cut-off,
 because of recombination to hydrogen atoms. A more detailed discussion
concernig the cloud temperature is given by Kang et al. (1990).

By using the above four equations (\ref{eqn:vir})---(\ref{eqn:ene})
 with three free parameters $F_{\rm{v}}, T_{\rm{c}}$, and $T_{\rm{h}}$,
we can obtain the equilibrium masses and radii of the hot region
and cold clouds
 as follows:
\begin{eqnarray}
 R_{\rm{h}} &=& 0.416\frac{1}{f} \times
           \frac{\Lambda(T_{\rm{h}})}{G(kT_{\rm{h}})^{0.5}m_{\rm{p}}^{1.5}} ,
\label{eqn:rh}\\
 M_{\rm{h}} &=& 3.41\frac{(1-F_{\rm{v}})}{f^{2}} \times
           \frac{\Lambda(T_{\rm{h}})(kT_{\rm{h}})^{0.5}}{G^{2}m_{
          \rm{p}}^{2.5}} ,
\label{eqn:mh}\\
 R_{\rm{c}} &=& 0.440\frac{F_{\rm{v}}^{2}}{f(1-F_{\rm{v}})} \times
          \frac{\Lambda(T_{\rm{h}})(\eta c^{2})}
               {Gm_{\rm{p}}^{0.5}k^{1.5}T_{\rm{h}}^{0.5}T_{\rm{c}}} ,
\label{eqn:rc}\\
 M_{\rm{c}} &=& 8.17\frac{F_{\rm{v}}^{6}}{f^{2}(1-F_{\rm{v}})^{3}} \times
           \frac{m_{\rm{p}}^{0.5}\Lambda(T_{\rm{h}})T_{\rm{h}}^{1.5}
            (\eta c^{2})^{3}}
           {G^{2}k^{2.5}T_{\rm{c}}^{4}}  ,
\label{eqn:mc}
\end{eqnarray}
where
\begin{eqnarray}
 f &=& (1-F_{\rm{v}})+2.02F_{\rm{v}}\frac{T_{\rm{h}}}{T_{\rm{c}}} .
\end{eqnarray}
\\

\it
2.2. Results
\rm

\baselineskip 8mm

In figure 1, we show the relation between the total mass
$M_{\rm{total}}(=M_{\rm{h}}+N_{\rm{c}}M_{\rm{c}})$ and the radius
$R_{\rm{h}}$
of a two-phase protogalaxy for $T_{\rm{c}}=10^{4}$ K, $\eta=10^{-6},$ and
five cases of $F_{\rm{v}}$. For a comparison,
we plot the case of a one-component
protogalaxy by a dashed line.
As is well-known, in the case of a one-component protogalaxy (no clouds)
the upper bound of the radius is obtained at the free-free cooling region.
On the other hand, in the case of a two-component protogalaxy the total mass
is limited lower with increasing the cloud fraction.
For example, the total mass
is bounded to $10^{13}M_{\odot}$ for $F_{\rm{v}}=10^{-3}$.
 For this result, we can give a clear physical interpretation.
The cooling time is determined by the hot gas, while the dynamical time
is controlled by clouds for a cloud-dominant galaxy.
That is, in a cloud-dominant galaxy the protogalaxy, itself, cannot
collapse, because of an inefficient energy loss from the hot region [see
 equation (\ref{eqn:tim})]. The upper mass bound which we derived here is
about two orders
larger than the typical mass of a galaxy, $\sim 10^{11} M_{\odot}$. We
discuss this point in section 4.

\vspace{1.0cm}

\large
\bf
{3.Stability of Clouds}
\normalsize
\rm
\vspace{0.5cm}

In this two-component model, we assume that cold clouds collide with each
other and liberate energy to the hot ambient medium.
  In order for this assumption to be satisfied, the cold clouds embedded
in the hot medium must be dynamically stable. Therefore,
we pose two stability conditions concerning the clouds.
\vspace{0.2cm}\\
\it
3.1. Gravitational Instability
\rm
\baselineskip 8mm

First, clouds must be gravitationally stable,
\begin{eqnarray}
 M_{c} {\rm{\leq}} M_{\rm{c,Jeans,p}},
\label{eqn:c1}
\end{eqnarray}
where the Jeans mass of a pressure-confined cloud, $M_{\rm{c,Jeans,p}}$,
 is given by (Spitzer 1978),
\begin{eqnarray}
 M_{\rm{Jeans,p}}=1.8\frac{c_{\rm{c}}^{4}}{G^{1.5}P_{\rm{h}}^{0.5}}.
\label{eqn:jp}
\end{eqnarray}
Here, $c_{\rm{c}}$ and $P_{\rm{h}}$ are the sound speed in the cloud
and the pressure of the
ambient hot medium, respectively. We apply (\ref{eqn:jp}) to the cold cloud,
\begin{eqnarray}
 M_{\rm{c,Jeans,p}}=1.38{\times}10^{7}{\times}\sqrt{1-F_{\rm{v}}}
              (\frac{T_{\rm{c}}}{10^{4}\rm{K}})^{2}(\frac{10^{6}\rm{K}}
              {T_{\rm{h}}})^{0.5}
              (\frac{R_{\rm{h}}}{100~\rm{kpc}})^{1.5}
               \sqrt{\frac{10^{13}M_{\odot}}{M_{\rm{h}}}}
\hspace{12pt}M_{\odot}.
\label{eqn:mcjj}
\end{eqnarray}
This mass is roughly comparable to  the mass of a globular cluster
(Fall, Rees 1985).
Of course, if we take $T_{\rm{c}}=10^{2}$ K, due to $\rm{H}_{2}$ cooling (
see Kang et al. 1990), this situation greatly changes.
In this case, according to equation
(\ref{eqn:mcjj}), the Jeans mass of a cold cloud is smaller by four orders
 of magnitude than that for the case $T_{\rm{c}}=10^{4}$ K.
\vspace{0.2cm}\\
\it
3.2. Kelvin-Helmholtz Instability
\rm

Second, clouds must be stable against the Kelvin-Helmholtz
instability (Chandrasekhar 1961; Murray et al. 1992). This instability occurs
 when two media having different densities are in relative motion.
  In the present situation, clouds with a density
$\rho_{\rm{c}}{\gg}\rho_{\rm{h}}$ move in
the hot medium with random velocity $\sigma_{\rm{v}}$.
This instability occur when the momentum from a cloud to hot gas is
transferred (cf. Chandrasekhar 1961), i.e.,
\begin{eqnarray}
 2\rho_{\rm{c}}R_{\rm{c}} &\geq& \rho_{\rm{h}}R_{\rm{h}}.
\end{eqnarray}
In another description, it reduces to
\begin{eqnarray}
 R_{\rm{c}}&\geq&R_{\rm{c,crit}}=\frac{1}{2}\frac{\rho_{\rm{h}}}
       {\rho_{\rm{c}}}R_{\rm{h}}
       =0.248\frac{T_{\rm{c}}}{T_{\rm{h}}}R_{\rm{h}}
\nonumber\\
       &=&2.48{\times}10^{2}(\frac{T_{\rm{c}}}
        {10^{4}~\rm{K}})(\frac{10^{6}~\rm{K}}
       {T_{\rm{h}}})
         \sqrt{\frac{R_{\rm{h}}}{100~\rm{kpc}}}{\hspace{12pt}}\rm{pc}.
\label{eqn:c2}
\end{eqnarray}
Since in this paper we assume that the cold clouds are confined by the
pressure of the hot ambient medium,
 the stabilizing mechanism against the Kelvin-Helmholz
instability, such as the self-gravity and magnetic field of the cold cloud,
can be neglected, and the above simple treatment for the clouds
is not sufficient.
However, this is the most generous criterion.
We call equations (\ref{eqn:c1}) and (\ref{eqn:c2})
the stability conditions, and $M_{\rm{c,Jeans,p}}$ and $R_{\rm{c,crit}}$
the critical mass and radius for the
cold clouds, respectively.
\vspace{0.2cm}\\
\it
3.3.Results
\rm
\baselineskip 8mm

In figure 2a, we show the mass of a cold cloud as a function of
the hot gas temperature $T_{\rm{h}}$ for $T_{\rm{c}}=10^{4}$ K,
 $\eta=10^{-6}$ and five cases
 of $F_{\rm{v}}$.
The solid and dashed lines show $M_{\rm{c}}$ and $M_{\rm{c,Jeans,p}}$,
which are calculated from equations (\ref{eqn:mc}) and (\ref{eqn:mcjj}),
 respectively.
In figure 2b, we show the radius of the cold cloud as a function of
the hot gas temperature for the same parameters as in figure 2a.
The solid and dashed lines show $R_{\rm{c}}$ and $R_{\rm{c,crit}}$,
which are calculated from equations (\ref{eqn:rc}) and (\ref{eqn:c2}),
respectively.
  For different $F_{\rm{v}}$, $M_{\rm{c,Jeans,p}}$
 and $R_{\rm{c,crit}}$  change only slightly
 while $M_{\rm{c}}$ sensitively changes.
The gravitational stability condition gives the upper limit of the cloud mass
(or $T_{\rm{h}}$), and the Kelvin-Helmholtz stability condition gives
 the lower
limit of the cloud radius (or $T_{\rm{h}}$). As can be seen in figure 2,
there are only small allowed regions of parameters for each $F_{\rm{v}}$.
Table 1 summerizes the allowed ranges of $T_{\rm{h}}$ for each model,
and table 2 gives the physical quantities of a two-component protogalaxy
when the median value of $T_{\rm{h}}$ from table 1 is taken for each model.
In our model, the cold clouds of mass ${\sim}10^{6}M_{\odot}$ are confined by
the thermal pressure of a hot ambient gas of mass ${\sim}10^{13}M_{\odot}$,
irrespective of $F_{\rm{v}}$.
\vspace{1.0cm}\\
\break
\large
\bf
4.Summary and Discussion
\rm
\normalsize
\baselineskip 8mm

{}~~~We considered the physical properties of a two-component
protogalaxy by using a simple analytic treatment,
and determined them considering the stability of the cold clouds.
In the cloud-dominant protogalaxy,
the mass of a protogalaxy is almost constant in the free-free cooling region
above ${\sim}10^{6}$ K.  This means that a protogalaxy
heavier than ${\sim}{10^{13}M_{\odot}}$ cannot collapse.
In contrast, the mass of an one-component protogalaxy is not limited.
The stability condition for cold clouds against the
Jeans instability and Kelvin
-Helmholtz instability poses stringent limitations on the physical
quantities of the two-component protogalaxy.

In our simple model, several important factors are neglected concerning the
formation of galaxies,
such as the abundance and distribution of dark matter
 (White, Rees 1978), the cosmological initial conditions,
the effect of shock heating (Thoul, Weinberg 1995),
the angular momentum and the deviation from spherical symmetry of a protogalaxy
 (De Araujo, Opher 1994).
To explain
the typical mass of a galaxy ($\sim 10^{11}M_{\odot}$), quantitatively,
 we must take into account the above-mentioned factors.
 The upper mass bound which we present here may give one possibility for the
 origin of the typical galaxy mass.

According to the present viewpoint of a two-component protogalaxy,
 we speculate on the star-formation history as follows.
In the protogalaxy with radius $R_{\rm{h}}$ ${\sim}$ $100$ kpc and a
total mass
of $M_{\rm{total}}$ ${\sim}$ $10^{13}M_{\odot}$, the star-formation
in an extended halo is triggered
by cloud-cloud collisions and/or the shock compression of clouds by the
 jet suggested in high-redshift radio galaxies
(McCarthy 1993; Norman, Ikeuchi 1994, private communication).
These Population II stars distribute over the extended halo,
 and their low mass
ends may be candidates for MACHOs.
Meanwhile, the hot gas is cooled and contracts to a disk or spheroid,
depending upon its angular momentum.
{}From this gas component,  Population I stars are formed.

In the succeeding work we consider the evolution of this two-component
protogalaxy.
 Especially, once the protogalaxy begins to collapse,
 the hot gas phase is cooled and contracts, while the cold clouds can not
contract,  because their random energy can not be extracted.
That is, the hot gas and cold clouds segregate.
 At this stage the stability of the cold clouds changes,
 because of the lack of hot ambient gas.
 Including these processes, we examine the evolution of two-component
protogalaxies.
\\

The authors would like to thank Professors M.Sasaki and N.Gouda for valuable
discussions, and
K.M. is grateful to Dr.M.Shibata for helpful suggestions and a
critical reading of the manuscript.\\
\newpage
\begin{center}
\bf{References}
\end{center}
\baselineskip 8mm
Binney J., Tremaine S. 1987, Galactic Dynamics (Princeton University
Press, Princeton) \\
Chandrasekhar S. 1961, Hydrodynamic and Hydromagnetic Stability
(Clarendon Press, Oxford). \\
Cen R., Ostriker J.P. 1993, ApJ 417, 404\\
De Araujo J.C.N., Opher R. 1994, ApJ 437, 556\\
Efstathiou G., Silk J. 1983,
Fund. Cosmic Phys. 9, 1\\
Fall S.M., Rees M.J. 1985, ApJ 298, 18\\
Gott J.R.III, Thuan T.X. 1976, ApJ 204, 649 \\
Ikeuchi S., Norman C.A. 1991, ApJ 375, 479 \\
Kang H., Shapiro P.R., Fall S.M., Rees M.J. 1990, ApJ 363, 488 \\
Lee S., Schramm D.N., Mathews G.J.  1995, ApJ 449, 616\\
McCarthy P.J. 1993, ARA$\&$A 31, 639  \\
Mathews G.J., Schramm D.N. 1993, ApJ 404,468 \\
Murray S.D., White S.D.M., Blondin J,M., Lin D.N.C. 1993, ApJ 407, 588 \\
Ostriker J.P. 1974, talk at 7th Texas Conference\\
Rees M.J., Ostriker J.P. 1977, MNRAS 179, 541\\
Rees M.J. 1978, Phys.Scripta 17, 371 \\
Silk J. 1977, ApJ  211, 638 \\
Spitzer L.Jr. 1978, Physical Processes in the Interstellar Medium
(Wiley Interscience,    New York).\\
Thoul A.N., Weinberg D.H. 1995, ApJ 442, 480\\
White S.D.M., Rees M.J. 1978, MNRAS 183, 341\\
\newpage
\begin{center}
Table 1.~Allowed values for four parameters with respect \\
to the Jeans and Kelvin-Helmholtz instability.
\end{center}
\vspace{12pt}
\begin{center}
\begin{tabular}{ccccc} \hline
Model& $\log{T_{\rm{c}}}$ & $\log{\eta}(10^{-6})$ &
$F_{\rm{v}}(10^{-4})$ &
$\log{T_{\rm{h}}}$(K)(allowed) \\
\hline
1 & & & 3.0 & 6.38---6.41 \\
2 & &
10 & 2.0 & 6.73---6.78 \\
3 &
\multicolumn{1}{c}{\raisebox{-2.0ex}[0pt]{4}} &
 & 1.0 & 7.33---7.40  \\
\cline{3-5}
4 & & & 10.0 & 6.34---6.42  \\
5 & & 1  & 8.0 & 6.53---6.64  \\
6 & & & 5.0 & 6.94---7.12 \\
\hline
\end{tabular}
\end{center}
\vspace{12pt}
\begin{center}
Table 2.~Physical quantities of a two-phase protogalaxy \\
for the median value of $T_{\rm{h}}$ in table 1.
\end{center}
\begin{center}
\vspace{12pt}
\begin{tabular}{cccccc} \hline
Model & $\log{T_{\rm{h}}}$(K) &
$\log{R_{\rm{h}}}$(pc) & $\log{M_{\rm{h}}}(M_{\odot})$ &
$\log{R_{\rm{c}}}$(pc) & $\log{M_{\rm{c}}}(M_{\odot})$ \\
\hline
1 & 6.40  & 5.39 & 12.93 & 2.41 & 6.68 \\
2 & 6.74  & 5.29 & 13.14 & 1.95 & 6.18 \\
3 & 7.36  & 5.13 & 13.51 & 1.18 & 5.37 \\
\cline{2-6}
4 & 6.40  & 5.28 & 12.70 & 2.34 & 6.59 \\
5 & 6.60  &5.19 & 12.77 & 2.06 &  6.28\\
6 & 7.00  &5.03 & 12.92 & 1.49 &  5.60 \\
\hline
\end{tabular}
\end{center}
\vspace{1.5cm}
\break
\large
\begin{center}
\bf{Figure Captions}
\end{center}
\normalsize
\bf{Fig. 1}\\
\rm
Relation between the total mass and radius of a two-component
protogalaxy for $T_{\rm{c}}=10^{4}$ K,
$\eta=10^{-6}$ and five cases of $F_{\rm{v}}$.
 For a comparison, the case without clouds is also plotted by a dashed line. \\
\bf{Fig. 2}\\
\rm
Cloud mass (a) and radius (b) as a function of
$T_{\rm{h}}$ for $T_{\rm{c}}=10^{4}$
K, $\eta=10^{-6}$ and five cases of $F_{\rm{v}}$.
The Jeans upper mass limit and
the Kelvin-Helmholtz lower radius limit are illustrated by dashed lines
 for the same parameters.
\newpage
}
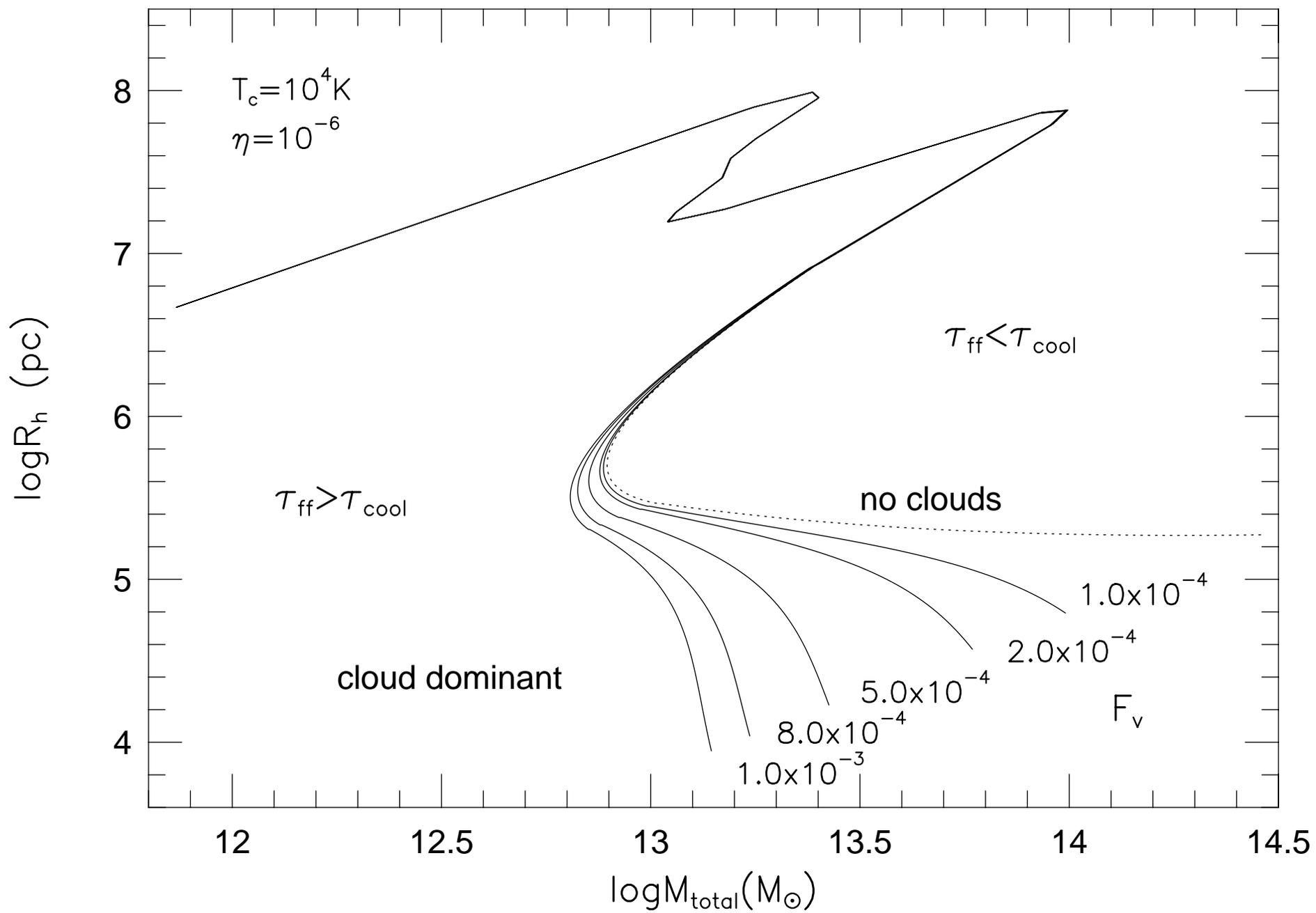
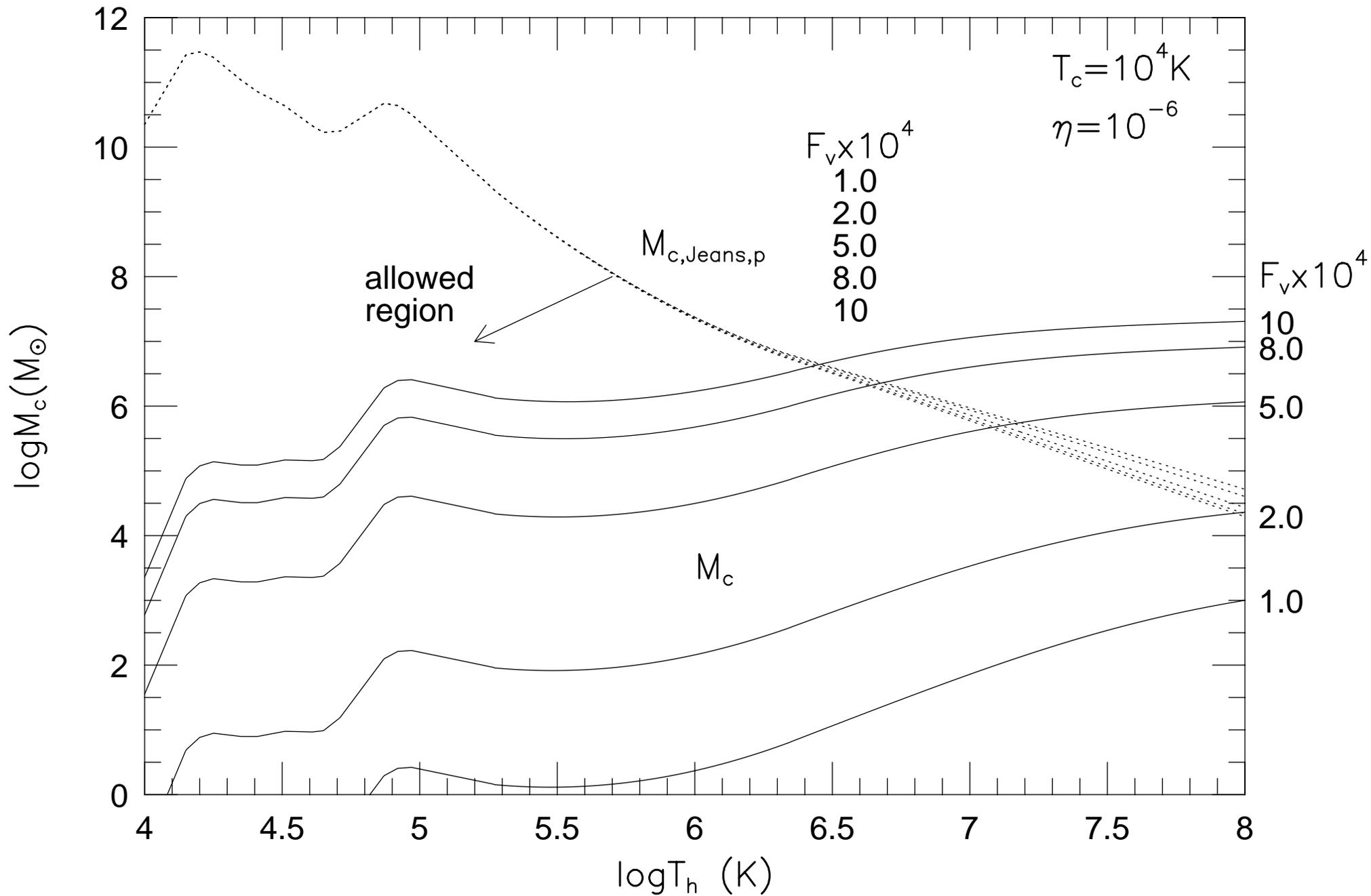
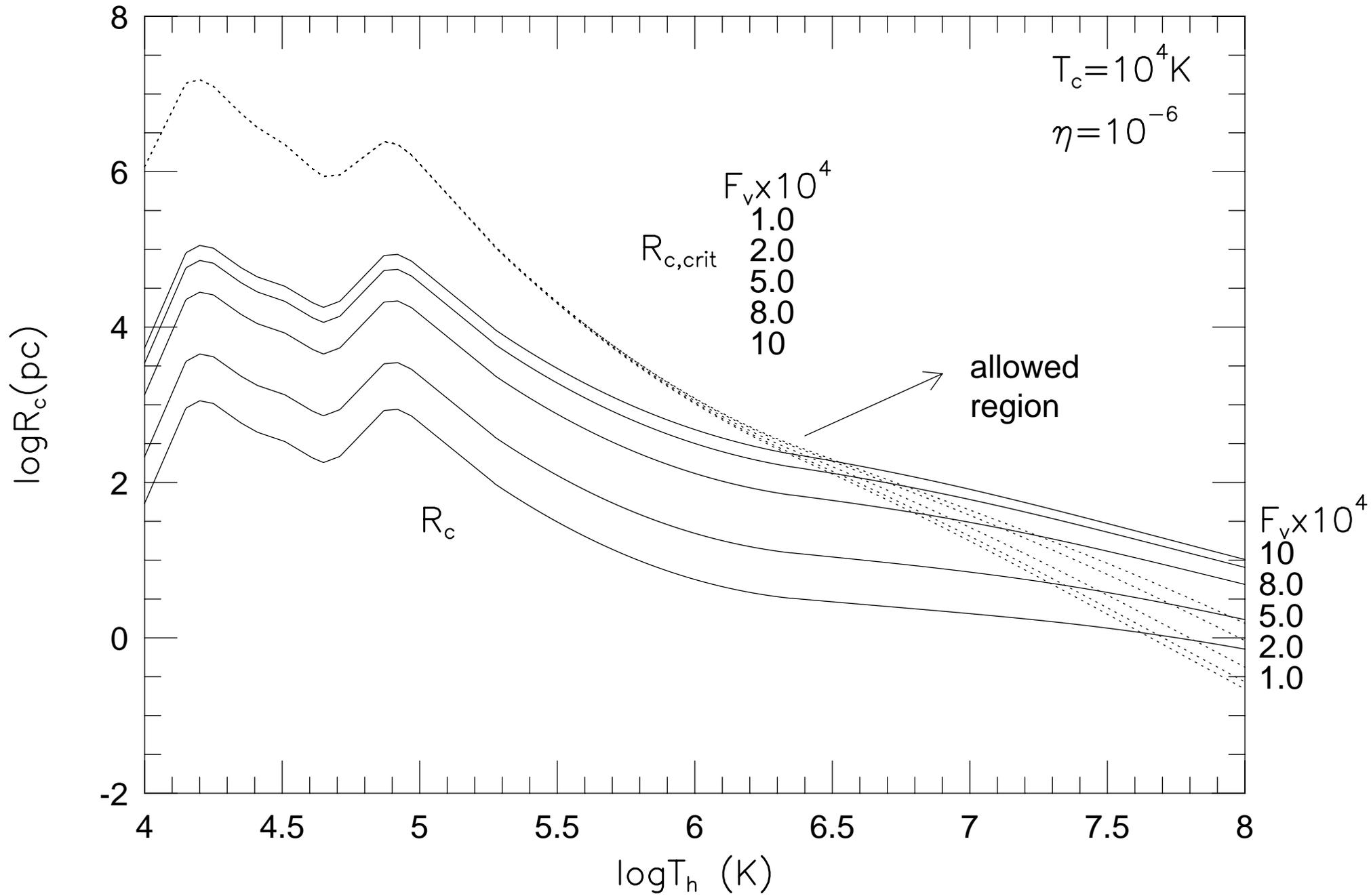
\end{document}